\documentclass
[superscriptaddress,secnumarabic,amssymb,amsmath,nobibnotes,aps,prd,showkeys,showpacs,nofootinbib]{revtex4}%
\usepackage{graphicx}
\usepackage{epsf}
\usepackage{bm}
\usepackage{amsmath}
\usepackage{amsfonts}
\usepackage{amssymb}
\usepackage{epstopdf}%
\providecommand{\U}[1]{\protect\rule{.1in}{.1in}}
\newcommand{\be}{\begin{equation}}
\newcommand{\ee}{\end{equation}}

\newcommand{\mincir}{\raise
-3.truept\hbox{\rlap{\hbox{$\sim$}}\raise4.truept\hbox{$<$}\ }}
\newcommand{\magcir}{\raise
-3.truept\hbox{\rlap{\hbox{$\sim$}}\raise4.truept\hbox{$>$}\ }}

\begin{document}
\title{f(R)-gravity from Killing Tensors}
\author{Andronikos Paliathanasis}
\email{anpaliat@phys.uoa.gr}
\affiliation{Instituto de Ciencias F\'{\i}sicas y Matem\'{a}ticas, Universidad Austral de
Chile, Valdivia, Chile}

\begin{abstract}
We consider $f\left(  R\right)  $-gravity in a
Friedmann-Lema\^{\i}tre-Robertson-Walker spacetime with zero spatial
curvature. We apply the Killing tensors of the minisuperspace in order to
specify the functional form of $f\left(  R\right)  $ and the field equations
to be invariant under Lie-B\"{a}cklund transformations which are linear in the
momentum (contact symmetries). Consequently, the field equations to admit
quadratic conservation laws given by Noether's Theorem. We find three new
integrable $f\left(  R\right)  $ models, for which with the application of the
conservation laws we reduce the field equations to a system of two first-order
ordinary differential equations. For each model we study the evolution of the
cosmological fluid. Where we find that for the one integrable model the
cosmological fluid has an equation of state parameter, in which in the latter
there is a linear behavior in terms of the scale factor which describes the
Chevallier, Polarski and Linder (CPL) parametric dark energy model.

\end{abstract}
\maketitle

\section{Introduction}

\label{intro}

The source for the late-time cosmic acceleration
\cite{Teg,Kowal,Komatsu,planck,Ade15} has been attributed to an unidentified
type of matter with a negative parameter in the equation of state, the dark
energy. The cosmological constant, $\Lambda$, leading to the $\Lambda
$-cosmology, is the simplest candidate for the dark energy. In $\Lambda
$-cosmology the universe consists of two perfect fluids, namely, the dust
fluid (dark matter) with zero pressure, and the dark energy fluid which
corresponds to the cosmological constant $\Lambda$, with parameter,
$w_{\Lambda}=-1$ in the equation of state. The terms which correspond to the
cosmological constant in the field equations can be seen in two ways, as a
cosmic fluid\ with constant energy density and constant negative parameter in
the equation of state, or as an additional component which follows from the
modification of the Einstein-Hilbert Action in General Relativity. However,
$\Lambda-$cosmology suffers from two major problems, the fine tuning and the
coincidence problems \cite{Weinberg1,Padmanabhan1,periv}.

In recent years other cosmological models have been introduced in order to
explain the acceleration phase of the universe. Some of them introduce a
cosmic fluid into Einstein's General Relativity
\cite{Ratra,Barrow,Linder,Copeland,Overduin,Basil,Kame,Bento,Supri}, while
some other models modify the Einstein-Hilbert Action
\cite{Brans,Buda,Sotiriou,Ferraro,Ferraro2,Maartens}.

In this work we are interested in the $f\left(  R\right)  $-gravity in a
Friedmann-Lema\^{\i}tre-Robertson-Walker spacetime (FLRW) with zero spatial
curvature. $f\left(  R\right)  $-gravity is a modified theory of gravity in
which the Action Integral of the field equations is a function $f$, of the
Ricci scalar $R$ of the underlying geometry \cite{Buda} (for review see
\cite{Sotiriou,odin1}). In the case for which $f\left(  R\right)  $ is a
linear function, $f\left(  R\right)  $-gravity reduces to standard General
Relativity, with or without the term due to the cosmological constant. The
functional form of $f\left(  R\right)  $ is still unknown and different forms
for $f\left(  R\right)  $ provide us with different dynamics, i.e. evolution
of the universe. The well-known Starobinsky model, with $f\left(  R\right)
=R+\alpha R^{2}$, has been proposed as an inflationary model of gravity
\cite{star}. A class of viable models which describe the accelerated expansion
of the universe can be found in \cite{odin2}. Other models which have been
proposed in the literature can be found in
\cite{fr1,fr2,fr3,fr5,fr6,fr7,fr7a,fr8,Ame10} and references cited therein,
while some cosmological data analysis of $f\left(  R\right)  $ models can be
found in \cite{data1,data2}

The purpose of this paper is the determination of the functional form of
$f\left(  R\right)  $ in order for the modified field equations to admit
quadratic conservation laws. To perform this analysis we use the method of
group invariant transformations, specifically the selection rule that we
assume is that the field equations are invariant under a contact
transformation which are defining as one-parameter transformations in the
tangent bundle of the dynamical system.

According to the well-known Noether's Theorems, the existence of a group
invariant transformation in the field equations is equivalent to the existence
of conservation flow. Noether's theorem states that for every one-parameter
transformation of the Action Integral of a Lagrangian function which
transforms the Action Integral in such a way that the Euler-Lagrange equations
are invariant. A conservation flow corresponds to the transformation. Contact
transformations are important in physical science as they are related with
important conservation laws such as the Runge-Lenz Vector, the Lewis
Invariant; contact transformations provide also the conservation law in the
MICZ-Problem \cite{micz1,micz2}

Group invariant transformations cover a range of applications in gravitational
physics and cosmology. For instance, Lie point symmetries have been used for
the determination of closed-form solutions in a model with charged perfect
fluids in spherically symmetric spacetimes \cite{Maharaj1,Maharaj3}. On the
other hand, Noether point symmetries have been introduced as a selection rule
for the determination of the functional form of the potential in scalar field
cosmology in \cite{deRitis}. Since then that method has been applied in
various cosmological models and new solutions have been found (for instance
see
\cite{Rosquist,Cap97M,CapP09,Cap3,KotsakisL,capPhantom,terzis,terzis2,vak2,vak3}
and references cited therein). In \cite{Basilakos1} Noether symmetries were
applied in the scalar field cosmological scenario as a geometric selection
rule for the determination of the functional form of the potential. In
theories with minisuperspace, Noether point symmetries are generated by the
collineations of the minisuperspace. Collineations are the generators of
one-parameter transformations which transform geometric objects under a
certain rule \cite{Yano}. However, the minisuperspace is defined by the
cosmological model and the existence of collineations depends upon the model.
Hence the requirement of the existence of a Noether point symmetry in the
cosmological Lagrangian is also a self-criterion because we let the theory
select the functional form of the model. That selection rule is consistent
with the geometric character of gravity. This geometric approach has been
applied in various cosmological models and new integrable cosmologically
viable models have arisen \cite{ref1,ref3,ref4}.

The application of Noether (point) symmetries which are the generators of
one-parameter point transformation for the $f\left(  R\right)  $-cosmology in
an FLRW spacetime, in which the spacetime comprises a perfect fluid with zero
pressure, has been performed before in \cite{palfr}. Recently the same
analysis has been performed and for a general perfect fluid with nonzero
equation of state parameter \cite{christ}, whereas for some locally rotational
spacetimes the Noether point symmetry classification of $f\left(  R\right)
$-cosmology can be found in \cite{ermakov}. Some other $f\left(  R\right)
$-models with closed-form solutions can be found in \cite{close1}, while the
application of point transformations in $f\left(  R\right)  $-gravity in
static spherically symmetric spacetimes can be found in \cite{capfs,capfs1}

Another geometric selection rule which is based upon the group invariant
transformations of the Wheeler-DeWitt equations was introduced in \cite{Abar}.
It has been shown that the existence of a Lie point symmetry for the
Wheeler-DeWitt equation is equivalent with the existence of a Noetherian
conservation law for the classical field equations. The main result was
applied in a scalar field cosmological model with a perfect fluid for which a
new integrable scalar field model was derived which it has been shown that the
model provide us with a viable inflationary scenario \cite{bbar}

The application of contact symmetries in cosmological studies is not new.
Contact symmetries have been applied for the determination of conservation
laws in various models in \cite{Rosquist}. In \cite{dynsym} contact symmetries
have been applied for the determination of the potential in scalar field
cosmology. As in the case of point symmetries, contact symmetries (or
Dynamical Noether symmetries) are also a geometric selection rule because they
follow from a class of collineations of the minisuperspace which are called
Killing tensors. Following the method which was presented in \cite{dynsym} we
apply the same selection rule for the determination of the unknown functional
form, $f~$, in modified $f\left(  R\right)  $-cosmology in the metric
formalism. The plan of the paper is as follows.

In Section \ref{field} we present the field equations in $f\left(  R\right)
$-gravity and we define our model which is a spatially flat FLRW spacetime and
contains a dustlike fluid which is not interacting with gravity in the Action
Integral. The basic properties of the Lie-B\"{a}cklund symmetries and
Noether's Theorem for contact transformations are given in Section
\ref{LBsym}.\ \ In Section \ref{conservationlaws}, we use the Killing tensors
of the minisuperspace in order to determine the $f\left(  R\right)  $-models
in which the field equations are invariant under contact transformations. For
each model we give the corresponding quadratic conservation law. We find five
models which admit quadratic conservation laws. Two of these have been found
from the application of Noether's Theorem for one-parameter point transformations.

In Section \ref{solutions} we apply the extra conservation laws in order to
reduce the order the dynamical system, which is defined by the field
equations. For the three new integrable models we show that two of the models
are supported by Lie surfaces while the third model is supported by a
Liouville surface. \ For each of the models we study the evolution of the
equation of state parameter for the cosmological fluid which corresponds to
$f\left(  R\right)  $-gravity. Appendix \ref{nonflat} completes our results
where we present the quadratic conservation laws of $f\left(  R\right)
$-gravity in a spatially nonflat FLRW spacetime. Finally in Section \ref{con},
we draw our conclusions.

\section{Field equations in $f\left(  R\right)  $-gravity}

\label{field}

We consider a FLRW universe with zero spatial curvature in which the
fundamental line element is given by the following expression%
\begin{equation}
ds^{2}=-dt^{2}+a^{2}\left(  t\right)  \left(  dx^{2}+dy^{2}+dz^{2}\right)  .
\label{fr.01}%
\end{equation}
The line element (\ref{fr.01}) describes an isotropic universe and admits as
Killing algebra the Killing vectors (KVs) of the three-dimensional Euclidean
space, that is, the $T_{3}\otimes_{s}SO\left(  3\right)  $ Lie algebra
consisting of the three translations and the three rotation symmetries of
$E^{3}$.

For the gravitation Action Integral we consider that of $f\left(  R\right)
$-gravity, that is,%
\begin{equation}
S=\int dx^{4}\sqrt{-g}\frac{1}{2k}f\left(  R\right)  +S_{m}, \label{fr.02}%
\end{equation}
where $S_{m}, =\int dx^{4}\sqrt{-g}L_{m}$, corresponds to the matter term, $R$
is the Ricci scalar of the underlying geometry and $k=8\pi G$. We assume
that$~S_{m}$ describes a dustlike fluid minimally coupled to gravity

Variation of the Action Integral (\ref{fr.02}) with respect to the metric
leads to the following field equations \cite{Sotiriou}%
\begin{equation}
f^{\prime}R_{\mu\nu}-\frac{1}{2}fg_{\mu\nu}-\left(  \nabla_{\mu}\nabla_{\nu
}-g_{\mu\nu}\nabla_{\sigma}\nabla^{\sigma}\right)  f^{\prime}=kT_{\mu\nu},
\label{fr.03}%
\end{equation}
where prime, $f^{\prime}\left(  R\right)  $, denotes total differentiation
with respect to $R,$~and $\nabla_{\mu}$ is the covariant derivative associated
with the Levi-Civita connection of the underlying Riemannian space with metric
tensor $g_{\mu\nu}$. Furthermore $R_{\mu\nu}$ is the Ricci tensor of
$g_{\mu\nu}$ and $T_{\mu\nu}$ is the energy-momentum tensor for the matter component.

Furthermore the energy-momentum tensor, $T_{\mu\nu}$, satisfies the Bianchi
identity $\nabla^{\nu}T_{\mu\nu}=0$. \ From the field equations (\ref{fr.03})
we observe that in the case for which $f\left(  R\right)  $ is a linear
function standard General Relativity is fully recovered.

In the context of the FLRW spacetime (\ref{fr.01}), in which the Ricci scalar
is\footnote{Overdot denotes total differentiation with respect to
\textquotedblleft$t$\textquotedblright.}
\begin{equation}
R=6\left(  \frac{\ddot{a}}{a}+\left(  \frac{\dot{a}}{a}\right)  ^{2}\right)
\label{fr.05}%
\end{equation}
and for a dustlike fluid with energy-momentum tensor $T_{\mu\nu}=\rho
_{m}u_{\mu}u_{\nu}$, where $u_{\mu}=\delta_{\mu}^{t},$ is the comoving
observer, $u^{\mu}u_{\mu}=-1$. Furthermore, from the Bianchi identity for the
tensor, $T_{\mu\nu}$, we derive the conservation law%
\begin{equation}
\dot{\rho}_{m}+3H\rho_{m}=0, \label{fr.08}%
\end{equation}
the solution of which is $\rho_{m}=\rho_{m0}a^{-3}$, where $\rho_{m0}$ is the
energy density for the dustlike fluid at the present time.

Hence from (\ref{fr.03}) we derive the following (modified) Friedmann's
equations%
\begin{equation}
3f^{\prime}H^{2}=k\rho_{m}+\frac{f^{\prime}R-f}{2}-3Hf^{\prime\prime}\dot{R},
\label{fr.06}%
\end{equation}
and%
\begin{equation}
2f^{\prime}\dot{H}+3f^{\prime}H^{2}=-2Hf^{\prime\prime}\dot{R}-\left(
f^{\prime\prime\prime}\dot{R}^{2}+f^{\prime\prime}\ddot{R}\right)
-\frac{f-Rf^{\prime}}{2}, \label{fr.07}%
\end{equation}
where $H=\dot{a}/a$ is the Hubble parameter.

Equations (\ref{fr.06}), (\ref{fr.07}) can be written as follows%
\begin{equation}
3H^{2}=k_{eff}\left(  \rho_{m}+\rho_{f}\right)  , \label{fr.09}%
\end{equation}
and%
\begin{equation}
2\dot{H}+3H^{2}=-k_{eff}p_{f}, \label{fr.10}%
\end{equation}
where $k_{eff}=k\left(  f^{\prime}\right)  ^{-1}$ is the effective
gravitational parameter and $\rho_{f},~p_{f}$ are the fluid components of
$f\left(  R\right)  $ gravity, that is,%
\begin{equation}
\rho_{f}=\frac{f^{\prime}R-f}{2}-3Hf^{\prime\prime}\dot{R} \label{fr.11}%
\end{equation}
and%
\begin{equation}
p_{f}=2Hf^{\prime\prime}\dot{R}+\left(  f^{\prime\prime\prime}\dot{R}%
^{2}+f^{\prime\prime}\ddot{R}\right)  +\frac{f-Rf^{\prime}}{2}. \label{fr.12}%
\end{equation}

Therefore the parameter of the equation of state (EoS) for the fluid
components of $f\left(  R\right)  $-gravity, $w_{f}=p_{f}/\rho_{f},$ has the
following expression%
\begin{equation}
w_{f}=\frac{p_{f}}{\rho_{f}}=-\frac{\left(  f-Rf^{\prime}\right)
+4Hf^{\prime\prime}\dot{R}+2\left(  f^{\prime\prime\prime}\dot{R}%
^{2}+f^{\prime\prime}\ddot{R}\right)  }{\left(  f-Rf^{\prime}\right)
+6Hf^{\prime\prime}\dot{R}},\label{fr.12a}%
\end{equation}
from where we can see that, when $f\left(  R\right)  =R-2\Lambda$, expression
(\ref{fr.12a}) gives $w_{f}=-1$. In order for $k_{eff}$ be a positive
function, $f^{\prime}>0,$ should hold. This is also required in order for the
final attractor of the field equations to be a de Sitter point (for details
see \cite{Ame10}). Furthermore, condition $f^{\prime}>0$ is necessary in order
to avoid the existence of ghosts . Another important constrain in which the
$f\left(  R\right)  $ function should satisfies is $f^{\prime\prime}>0$, for
$R\geq R_{0}$, where $R_{0}$, is the Ricci scalar today, in order the theory
to be consistency with local gravity tests. On the other hand, the violation
of the later constrain introduce tachyonic instability, and a nonwell defined
post-Newtonian limit \cite{frfelice,Olmo}. Furthermore, from the solar system
tests we have that $f(R)\approx R-2\Lambda,$ which means that $f\left(
R\right)  $ should reduced to General Relativity. Finally, if we assume that
the at the late-time the model has a stable de Sitter behavior then the
following condition should be satisfied, $0<\frac{Rf^{^{\prime\prime}}%
}{f^{^{\prime}}}(r)<1$ at $r=-\frac{Rf^{^{\prime}}}{f}=-2$ \ \cite{Ame10}. 

\subsection{Lagrange multiplier and minisuperspace}

In contrast to General Relativity, which is a second-order theory, $f\left(
R\right)  $-gravity is a fourth-order theory. This can be seen by substituting
the Ricci scalar (\ref{fr.05}) into (\ref{fr.07}). Another way to derive the
modified Friedmann's equations (\ref{fr.06}), (\ref{fr.07}) and (\ref{fr.05}),
is with the use of a Lagrange multiplier\footnote{For applications of the
Lagrange multipliers in high-order theories of gravity see
\cite{lan2,lan3,lan4,lan5}.} \cite{lan1}. \ Lagrange multipliers are useful to
reduce the order of the differential equations. However, at the same time the
dimension of the space of the dependent variables is increased. \ 

With the use of a Lagrange multiplier $\lambda,$ in the gravitational action
integral (\ref{fr.02}), we have%
\begin{equation}
S=\frac{1}{2k}\int dx^{4}\sqrt{-g}\left[  f\left(  R\right)  -\lambda\left(
R-6\left(  \frac{\ddot{a}}{a}+\left(  \frac{\dot{a}}{a}\right)  ^{2}\right)
\right)  \right]  +\int dx^{4}\sqrt{-g}\rho_{m0}a^{-3}%
\end{equation}
where we have used equation (\ref{fr.05}) and the solution of the Bianchi
identity (\ref{fr.08}) for the perfect fluid. Furthermore, the condition
$\frac{\partial S}{\partial R}=0$, gives $\lambda=f^{\prime}\left(  R\right)
$.

Hence we find that the Lagrangian of the modified Friedmann's equations is%
\begin{equation}
L\left(  a,\dot{a},R,\dot{R}\right)  =6af^{\prime}\dot{a}^{2}+6a^{2}%
f^{\prime\prime}\dot{a}\dot{R}+a^{3}\left(  f^{\prime}R-f\right)  +\rho_{m0}.
\label{fr.13}%
\end{equation}

Therefore, the field equations can be seen as the Euler-Lagrange equations of
(\ref{fr.13}) with respect to the variables $\left\{  a,R\right\}  $ and the
first modified Friedmann's equation (\ref{fr.06}) can be seen as the
Hamiltonian constraint\footnote{Alternatively equation (\ref{fr.06}) can be
derived from the Euler-Lagrange equation with respect to a new variable, $N,$
which arises from the lapsed time $dt=Nd\tau$ \cite{Ray}.} of the dynamical
system with Lagrangian (\ref{fr.13}), that is,
\begin{equation}
\mathcal{E}=6af^{\prime}\dot{a}^{2}+6a^{2}f^{\prime\prime}\dot{a}\dot{R}%
-a^{3}\left(  f^{\prime}R-f\right)  , \label{fr.14}%
\end{equation}
where the constant $\mathcal{E}$ is related to $\rho_{m0}$ as follows,
$\mathcal{E}=2k\rho_{m0}$ or $\mathcal{E}=6\Omega_{m}H_{0}^{2}$. In the last
expression $\Omega_{m}=k\rho_{m0}/\left(  3H_{0}^{2}\right)  $ and $H_{0}$ is
the Hubble constant.

We observe that Lagrangian (\ref{fr.13}), describes the motion of a particle
in the space of variables $\left\{  a,R\right\}  $, and it is in the form
$L=K+U$, where $K$ is the Kinetic energy which defines the minisuperspace with
line element%
\begin{equation}
ds^{2}=12af^{\prime}da^{2}+6a^{2}f^{\prime\prime}dadR, \label{fr.14a}%
\end{equation}
and $U=a^{3}\left(  f^{\prime}R-f\right)  +\rho_{m0}$, is the effective potential.

On the other hand, the field equations (\ref{fr.03}) describe the evolution of
the scale factor $a\left(  t\right)  $ in a fourth-order theory. The latter
means that the introduction of the Lagrange multiplier $\lambda$, in the
gravitational action integral, reduced the order of the theory to that of a
second-order theory whereas in the same time the degrees of freedom increased.
For an extensive discussion on the degrees of freedom in modified theories of
gravity see \cite{sott}.

In the following sections we discuss the application of contact
transformations of differential equations. We apply as a selection rule for
the determination of the unknown function $f\left(  R\right)  $ the
requirement that the Action Integral of the field equations with Lagrangian
(\ref{fr.13}) be invariant under contact transformations, equivalently the
dynamical system to admit dynamical Noether symmetries, consequently,
quadratic conservation laws. Below we assume that $f^{\prime\prime}\left(
R\right)  \neq0,$ otherwise Lagrangian (\ref{fr.13}) is that of General Relativity.

\section{Killing Tensors and Noether's Theorem}

\label{LBsym}

By definition a vector field $X$ is called a Lie-B\"{a}cklund symmetry of a
second-order differential equation $\Xi\left(  t,x^{k},\dot{x}^{k},\ddot
{x}^{k}\right)  =0,$ when $X$ is the generator of the infinitesimal
transformation \cite{Ibrag,Ibrag2},%
\begin{align}
t^{\prime}  &  =t+\varepsilon\xi\left(  t,x^{k},\dot{x}^{k}\right)
,\label{tr.01}\\
x^{\prime i}  &  =x^{i}+\varepsilon\eta^{i}\left(  t,x^{k},\dot{x}^{k}\right)
, \label{tr.02}%
\end{align}
that is, $X=\frac{\partial t^{\prime}}{\partial\varepsilon}\partial_{t}%
+\frac{\partial x^{\prime i}}{\partial\varepsilon}\partial_{i}$, which leaves
invariant the differential equation $\Xi$, i.e., $\Xi\left(  t^{\prime
},x^{\prime k},\dot{x}^{\prime k},\ddot{x}^{\prime k}\right)  =0,$ or
$X^{\left[  2\right]  }\left(  \Xi\right)  =0$, where $X^{\left[  2\right]  }$
is the second prolongation of $X,~$\cite{Bluman}.

Infinitesimal transformations of special interest are the (Lie) point
transformations, i.e., $\frac{\partial\xi}{\partial\dot{x}^{k}}=\frac
{\partial\eta^{i}}{\partial\dot{x}^{k}}=0$, and the contact transformations in
which $\xi,\eta$ are linear functions of the first derivatives of the
dependent variables. For the Lie-B\"{a}cklund transformations (except the
point transformations) it has been shown that transformation (\ref{tr.01}),
(\ref{tr.02}), is equivalent to the transformation \cite{Bluman,Sarlet}%
\begin{equation}
x^{\prime i}=x^{i}+\varepsilon\zeta\left(  t,x^{k},\dot{x}^{k}\right)  ,
\label{tr.03}%
\end{equation}
with generator $\bar{X}=\zeta\left(  t,x^{k},\dot{x}^{k}\right)  \partial_{i}%
$. Transformation (\ref{tr.03}) is called the canonical transformation of
(\ref{tr.01}), (\ref{tr.02}) and $\bar{X}$ is the canonical form of $X$.

In this work we are interested in contact transformations for which the
generator, $\bar{X}$, has the following form
\begin{equation}
\bar{X}=K_{j}^{i}\left(  t,x^{k}\right)  \dot{x}^{j}\partial_{i},
\label{tr.04}%
\end{equation}
where the second-rank tensor $K_{ij}$ is symmetric on the indices, that is,
$K_{\left[  ij\right]  }=0$.

For differential equations which arise from a variational principle there are
the two well-known Noether's Theorems \cite{EmmyN}. The first theorem relates
the action of the transformation (\ref{tr.01}), (\ref{tr.02}) on the Action
Integral for the Lagrangian $L=L\left(  t,x^{k},\dot{x}^{k}\right)  $ of the
differential equations, $\Xi$, in order for the latter to be invariant.
Specifically, if there exists a function $\sigma$ such that
\begin{equation}
X^{\left[  1\right]  }L+L\dot{\xi}=\dot{\sigma}, \label{tr.05}%
\end{equation}
then the Euler-Lagrange equations of Lagrangian $L$ are invariant under the
action of the transformation (\ref{tr.01}), (\ref{tr.02}) and~the generator
$X$ is called a Noether symmetry. $X^{\left[  1\right]  },$ denotes the first
prolongation of $X$.

Condition (\ref{tr.05}) is that which was originally introduced by Emmy
Noether in her originally paper \cite{EmmyN}. Incorrectly it is termed as
Noether Gauge symmetry condition \cite{NGS1,NGS2,NGS3}. Function $\sigma$ is
not a gauge function, but a boundary term introduced to allow for the
infinitesimal changes in the value of the Action Integral produced by the
infinitesimal change in the boundary of the domain caused by the infinitesimal
transformation of the variables in the action integral.

The second Noether's Theorem relates the existence of Noether symmetries with
that of conservation laws. Hence, if $X$ is the generator of the infinitesimal
transformation (\ref{tr.01}), (\ref{tr.02}) which satisfies the symmetry
condition (\ref{tr.05}) for a specific function$~\sigma$, then the function%
\begin{equation}
I=\xi\left(  \frac{\partial L}{\partial\dot{x}^{k}}\dot{x}^{k}-L\right)
-\eta^{i}\frac{\partial L}{\partial\dot{x}^{i}}+\sigma, \label{tr.07}%
\end{equation}
is a conservation law for the dynamical system with Lagrangian $L$. \ However,
for nonpoint transformations we can always apply the canonical transformation
(\ref{tr.04}) and condition (\ref{tr.05}) takes the simple form
\begin{equation}
\bar{X}^{\left[  1\right]  }L=\dot{\sigma}.
\end{equation}

As we discussed above, Lagrangian (\ref{fr.13}), describes an autonomous
dynamical system and it is in the form
\begin{equation}
L\left(  x^{i},x^{j}\right)  =\frac{1}{2}\gamma_{ij}\dot{x}^{i}\dot{x}%
^{j}+V_{eff}\left(  x^{k}\right)  , \label{tr.08}%
\end{equation}
where, $\gamma_{ij},$ is the minisuperspace of the field equations with line
element (\ref{fr.14a}), or equivalently,%
\begin{equation}
ds_{\left(  \gamma\right)  }^{2}=12a\phi da+12a^{2}dad\phi, \label{tr.10}%
\end{equation}
and effective potential, $V_{eff}\left(  x^{k}\right)  =a^{3}V\left(
\phi\right)  +\rho_{m0}$, where now the new variable $\phi,$ is $\phi
=f^{\prime}\left(  R\right)  ,$ and
\begin{equation}
V\left(  \phi\right)  =\left(  f^{\prime}R-f\right)  . \label{tr.11}%
\end{equation}
We remark that in the coordinates $\left\{  a,\phi\right\}  $ the Lagrangian,
(\ref{fr.13}), is that of a Brans-Dicke scalar field with zero Brans-Dicke
parameter \cite{Sotiriou,frfelice} which is also called O'Hanlon massive
dilaton gravity \cite{Hanlon}. Note that equation (\ref{tr.11}) is the
first-order Clairaut equation.

For the space of the dependent variables $\left\{  a,\phi\right\}  $, we have
that the general form of the contact symmetry (\ref{tr.04}) has the following
form%
\begin{align}
\bar{X}  &  =\left(  K_{~~a}^{a}\left(  t,a,\phi\right)  \dot{a}+K_{~\phi}%
^{a}\left(  t,a,\phi\right)  \dot{\phi}\right)  \partial_{a}+\label{tr.12}\\
&  ~~\ ~+\left(  K_{~a}^{\phi}\left(  t,a,\phi\right)  \dot{a}+K_{~\phi}%
^{\phi}\left(  t,a,\phi\right)  \dot{\phi}\right)  \partial_{\phi}\nonumber
\end{align}
\newline

The contact transformations with generator (\ref{tr.04}), which are Noether
symmetries\footnote{Usually these transformations are called dynamical Noether
symmetries.} for Lagrange functions of the form (\ref{tr.08}), have been
studied previously in the literature and it has been shown that the tensor
field $K_{j}^{i}$ of (\ref{tr.04}) is time-independent and is a Killing Tensor
for the metric $\gamma_{ij}$, that is, $\left[  \mathbf{K},\mathbf{\gamma
}\right]  _{SN}=K_{\left(  ij;k\right)  }=0$, where $\left[  ,\right]  _{SN},$
denotes the Schouten--Nijenhuis Bracket\footnote{For some applications of the
Killing tensors in General Relativity see for instance \cite{KT1,KT2,KT3}.}
and the following condition holds,
\begin{equation}
K_{(i}^{~~~~j}V_{_{eff},j)}+\sigma_{,i}=0, \label{tr.13}%
\end{equation}
in which $\sigma=\sigma\left(  x^{k}\right)  $,
\cite{Crampin1,Crampin,Kalotas}. Furthermore, from (\ref{tr.07}) it follows
that the corresponding Noether conservation law (\ref{tr.07}), is
time-independent and quadratic in the momentum.

For the space (\ref{tr.10}), the vector field (\ref{tr.12}) is
time-independent, that is, $\left(  K_{~a}^{a}\right)  _{,t}=\left(  K_{~\phi
}^{a}\right)  _{,t}=\left(  K_{~\phi}^{\phi}\right)  _{,t}=0$, and $K_{ij}$,
is a Killing Tensor of the minisuperspace (\ref{tr.10}). Furthermore condition
(\ref{tr.13}) gives the following system%
\begin{equation}
K_{~a}^{a}\left(  aV_{,\phi}\right)  +\left(  3V_{\phi}-2\phi V_{,\phi
}\right)  K_{~\phi}^{a}+a^{-2}\sigma_{\phi}=0, \label{tr.14}%
\end{equation}%
\begin{equation}
V_{,\phi}\left(  aK_{~a}^{\phi}-2\phi K_{~\phi}^{\phi}\right)  +3VK_{~\phi
}^{\phi}+a^{-2}\sigma_{,a}-2a^{-3}\phi\sigma_{,\phi}=0. \label{tr.15}%
\end{equation}

Hence in order to solve the system (\ref{tr.14}), (\ref{tr.15}) and determine
the functions $V\left(  \phi\right)  ,$ in which the modified field equations
admit quadratic conservation laws, we have to find the Killing Tensors
(KTs)~$K_{ij}$, of the two-dimensional metric (\ref{tr.10}). It is easy to see
that the Ricci scalar of (\ref{tr.10}) vanishes which means that $\gamma_{ij}$
is a flat space. However, the signature of $\gamma_{ij}$ is Lorentzian, that
is, the Lagrangian, (\ref{tr.13}), describes the motion of a particle in the
$M^{2}$ space. Moreover, because $\gamma_{ij}$ is the flat space, the Killing
Tensors are reducible \cite{Weir,barnesKT}. The latter gives that the KTs are
constructed by the tensor product of the Killing vectors (KVs) of $\gamma
_{ij}$. The Killing vectors and the Killing tensors of the two-dimensional
space (\ref{tr.11}) are given in Appendix \ref{killing}.

In the following we give the form of the potential $V\left(  \phi\right)  $,
the corresponding function $f\left(  R\right)  ,$ and the quadratic
conservation law which follow from the symmetry conditions (\ref{tr.14}),
(\ref{tr.15}).

\section{$f\left(  R\right)  $-gravity with quadratic conservation laws}

\label{conservationlaws}

The requirement of the existence of group invariant transformations,
symmetries, on theories of gravity is twofold; symmetries can be used as a
geometric selection rule to constrain the unknown parameters of the models and
to derive new analytical solutions. Furthermore this selection rule is
consistent with the geometric character of gravitational theories. The reason
is that the group invariant transformation of the field equations is related
with the geometry which the dependent variables define, in our consideration
with the minisuperspace (\ref{tr.10}). There exists a unique connection
between the collineations of the minisuperspace and the symmetries of the
field equations. In particular Noether point symmetries are related with the
elements of the conformal algebra of the minisuperspace \cite{Basilakos1},
whereas as we discussed before contact symmetries are generated by the KTs of
the minisuperspace. Hence, in such a way, by using as a selection rule the
existence of group invariant transformation in the field equations we let the
theory select the corresponding model.

On the other hand, analytical solutions are important in order to understand
the evolution of the universe. For instance, the field equations in $\Lambda
$-cosmology are maximally symmetric and are invariant under the same group of
invariant transformations with that of the linear second-order differential
equation\footnote{Specifically the dynamical system of $\Lambda$-cosmology is
that of the \textquotedblleft hyperbolic oscillator\textquotedblright.}.
Therefore symmetries can be used in order to recognize, or define, well-known
systems in gravity.

In our consideration the solution of the system (\ref{tr.14}), (\ref{tr.15}),
provides us with the function form of $V\left(  \phi\right)  $, consequently
the $f\left(  R\right)  $-theory, in which the field equations admit quadratic
conservation laws. The existence of a conservation law, which is in involution
and independent with the Hamiltonian, for the field equations which Lagrangian
(\ref{fr.13}) defines the evolution of the system in the phase-space, that is,
the field equations form an integrable dynamical system.

From the system (\ref{tr.14}), (\ref{tr.15}), we find that the only KT which
produces a contact symmetry for arbitrary function $V\left(  \phi\right)  $,
is the metric tensor $\gamma_{ij}$. The contact symmetry is the Hamiltonian
flow and the corresponding Noetherian conservation law is the Hamiltonian
(\ref{fr.14}). However, for specific $V\left(  \phi\right)  $, i.e. $f\left(
R\right)  $ function, the Lagrangian (\ref{fr.13}), is invariant under
additional contact symmetries. We have the following cases\footnote{Recall
that we consider $f^{\prime\prime}\left(  R\right)  \neq0$, that is,
$V_{,\phi\phi}\neq0$.}.

(I) When $V_{I}\left(  \phi\right)  =V_{1}\phi+V_{2}\phi^{3}$, the field
equations admit the quadratic conservation law%
\begin{equation}
I_{I}=3\left(  \phi\dot{a}+a\dot{\phi}\right)  ^{2}-V_{1}~a^{2}\phi^{2}
\label{fr.15}%
\end{equation}
generated by the KT $K_{22}^{ij}$.

(II) If $V_{II}\left(  \phi\right)  =V_{1}\phi-V_{2}\phi^{-7},~$ the field
equations admit the Noetherian conservation law%
\begin{equation}
I_{II}=3a^{4}\left(  \phi\dot{a}-a\dot{\phi}\right)  ^{2}+4V_{2}a^{6}\phi
^{-6}, \label{fr.16}%
\end{equation}
which follows from$~K_{11}^{ij}$.

(III)\ For $V_{III}\left(  \phi\right)  =V_{1}-V_{2}\phi^{-\frac{1}{2}},~$the
KT $K_{13}^{ij}$ generates a contact transformation for the field equations in
which the corresponding conservation law is
\begin{equation}
I_{III}=6a^{3}\dot{a}\left(  a\dot{\phi}-\phi\dot{a}\right)  -a^{5}\left(
\frac{3}{5}V_{1}-V_{2}\phi^{-\frac{1}{2}}\right)  . \label{fr.17}%
\end{equation}

(IV) When $V_{IV}\left(  \phi\right)  =V_{1}\phi^{3}+V_{2}\phi^{4},$ the
contact symmetry of the field equations follows from the KT $K_{12}^{ij}$,
which produce the conservation law%
\begin{equation}
I_{IV}=12a^{2}\left(  a^{2}\dot{\phi}^{2}-\phi^{2}\dot{a}^{2}\right)  +\left(
a\phi\right)  ^{4}\left(  3V_{1}+4V_{2}\phi\right)  . \label{fr.18}%
\end{equation}

(V) Finally, for $V_{V}\left(  \phi\right)  =V_{1}\left(  \phi^{3}+\beta
\phi\right)  +V_{2}\left(  \phi^{4}+6\beta\phi^{2}+\beta^{2}\right)  ,~$the
field equations admit the Noetherian conservation law
\begin{align}
I_{V}  &  =12a^{2}\left[  \left(  \beta-\phi^{2}\right)  \dot{a}^{2}+a^{2}%
\dot{\phi}^{2}\right]  +\nonumber\\
&  -a^{4}\left(  \beta-\phi^{2}\right)  \left[  V_{1}\left(  \beta+3\phi
^{2}\right)  +4V_{2}\left(  3\beta\phi+\phi^{3}\right)  \right]  ,
\label{fr.19}%
\end{align}
which follows from the linear combination of the two Killing tensors,
$K_{12}^{ij}+\beta K_{33}^{ij}$.

We continue with the solution of the Clairaut equation, (\ref{tr.11}), which
provides us with the corresponding $f\left(  R\right)  $ functions.

\subsection{$f\left(  R\right)  $-models}

In order to determine the form of function $f\left(  R\right)  ,$ from the
potentials $V_{I}-V_{V}$ we have to solve the Clairaut equation,
(\ref{tr.11}). The Clairaut equation always admits the linear solution,
$f\left(  R\right)  =f_{0}R+f_{1},$ and a singular solution is given by the
differential equation, $V_{,\phi}-R=0$ \cite{clairaut}. In our consideration
we are interested in the singular solution which provides us with the
functional form of $f\left(  R\right)  $ for each potential.

Hence, for $V_{I}\left(  \phi\right)  $, we find that the corresponding
$f\left(  R\right)  $ function is
\begin{equation}
f_{I}\left(  R\right)  =f_{0}^{I}\left(  R-V_{1}\right)  ^{\frac{3}{2}},
\label{fr.20}%
\end{equation}
in which $f_{0}=\frac{2\sqrt{3}}{9\sqrt{V_{2}}}.~$

Moreover from $V_{II}\left(  \phi\right)  $ we have
\begin{equation}
f_{II}\left(  R\right)  =f_{0}^{II}\left(  R-V_{1}\right)  ^{\frac{7}{8}},
\label{fr.21}%
\end{equation}
where $f_{0}^{II}=\frac{8}{7}\left(  7V_{2}\right)  ^{\frac{7}{8}}$.

For $V_{III}\left(  \phi\right)  $ the corresponding $f\left(  R\right)  $
function is
\begin{equation}
f_{III}\left(  R\right)  =f_{0}^{III}R^{\frac{1}{3}}-V_{1}, \label{fr.22}%
\end{equation}
where $f_{0}^{III}=3\left(  2^{-1}V_{2}\right)  ^{\frac{2}{3}}$. \ 

The $f\left(  R\right)  $ function which corresponds to the potential
$V_{IV}\left(  \phi\right)  $ is given by
\begin{equation}
f_{IV}\left(  R\right)  =\frac{1}{4V_{2}}\int\left[  \frac{\left(  F\left(
R\right)  \right)  ^{2}+V_{1}^{2}}{F\left(  R\right)  }-V_{1}\right]  dR,
\label{fr.23}%
\end{equation}
where~%
\begin{equation}
\left(  F\left(  R\right)  \right)  ^{3}=8V_{2}^{2}R-V_{1}^{3}+4V_{2}%
\sqrt{4V_{2}^{2}R^{2}-RV_{1}^{3}}. \label{fr.23a}%
\end{equation}

For $V_{2}=0$, (\ref{fr.23}) gives $f_{IV}^{1}\left(  R\right)  =\frac
{2\sqrt{3}}{9\sqrt{V_{2}}}R^{\frac{3}{2}}$, whereas for$~V_{1}=0$ we have the
closed-form solution, $f_{IV}^{2}\left(  R\right)  =\frac{3}{8}\left(
2V_{2}^{-1}\right)  ^{1/3}R^{4}$. Furthermore from (\ref{fr.23a}) we have that
$\operatorname{Im}F^{3}\left(  R\right)  =0$ when $R\left(  4V_{2}^{2}%
R-V_{1}^{3}\right)  >0$. Moreover, for $R>>V_{1}^{3}/V_{2}^{2},$ we have that
$F\left(  R\right)  \simeq g_{1}R^{1/3}$. Hence from (\ref{fr.23}) we find
that
\begin{equation}
f_{IV}^{3}\left(  R\right)  \simeq\frac{V_{1}}{4V_{2}}R+\frac{3}{16V_{2}g_{1}%
}\left(  2V_{1}\left(  g_{1}R\right)  ^{2/3}+\left(  g_{1}R\right)
^{4/3}\right)  . \label{fr.23b}%
\end{equation}
However, for $R<<V_{1}^{3}/V_{2}^{2}$, $F\left(  R\right)  $ is constant which
gives the limit of General Relativity.

Finally from $V_{V}\left(  \phi\right)  $ we have that
\begin{equation}
f_{V}\left(  R\right)  =\frac{1}{4V_{2}}\int\left[  \frac{\left(  \bar
{F}\left(  R\right)  \right)  ^{2}+\Sigma_{0}}{\bar{F}\left(  R\right)
}-V_{1}\right]  dR. \label{fr.24}%
\end{equation}
In the last expression $\Sigma_{0}=\left(  V_{1}^{2}-16\beta V_{2}^{2}\right)
~$and~the function $\bar{F}\left(  R\right)  $ is given by
\begin{align}
\left(  \bar{F}\left(  R\right)  \right)  ^{2}  &  =8V_{2}^{2}R-V_{1}%
\Sigma_{0}+\nonumber\\
&  +4V_{2}\sqrt{4V_{2}^{2}R^{2}-V_{1}R\Sigma_{0}+\beta\Sigma_{0}^{2}}.
\label{fr.25}%
\end{align}
We can see that again, for $R>>\left(  V_{1}^{2}-16\beta V_{2}^{2}\right)
/V_{2}^{2}$, function $f_{V}\left(  R\right)  $ is given by (\ref{fr.23b}).

In the case for which $\Sigma_{0}=0,~$i.e., $V_{1}=\pm4V_{2}\sqrt{\beta}$, the
potential $V_{V}\left(  \phi\right)  $ becomes $V_{V}\left(  \phi\right)
\simeq\left(  \phi\pm\sqrt{\beta}\right)  ^{4},$ whereas from expressions
(\ref{fr.24}) and (\ref{tr.11}) we derive that%
\begin{equation}
f_{V}^{1}\left(  R\right)  =\mp\sqrt{\beta}R+f_{1}R^{\frac{4}{3}},
\label{fr.26}%
\end{equation}
in which the new constant is $f_{1}=3\left(  2^{-1}V_{2}\right)  ^{\frac{8}%
{3}}$.

Furthermore, the quadratic conservation laws, $I_{I-V}$, are independent with
the Hamiltonian (\ref{fr.14}) and it holds that $\left\{  I_{I-V}%
,\mathcal{E}\right\}  =0$, where $\left\{  ,\right\}  $ is the Poisson
Bracket. Hence the models, $f_{I-V}$, are integrable.

In section \ref{field} we discussed some conditions in which the $f\left(
R\right)  $-theory should satisfy in order to be viable. It is easy to see
that the analytical $f\left(  R\right)  $-models, $f_{I}$,~$f_{II}$,~$f_{III}%
$,~$f_{IV}^{3}$ and $f_{V}^{1}$, satisfy the condition $f^{\prime}>0$, which
indicates that the theories are ghost free. As far as concerns the stability
condition $f^{\prime\prime}>0$, the models $f_{II}$, and $f_{III}$ violate the
condition, hence tachyonic instability presented.

However another important constrain on the $f\left(  R\right)  $-models is
that at the limit $R\rightarrow0$, the theory should has a similar behavior
with that of General Relativity. From the above models only~the $f_{V}%
^{1}\left(  R\right)  $ model gives that $f\left(  R\rightarrow0\right)
\simeq R$, while for the $f_{IV}^{3}$ model holds $f_{IV}^{3}\left(
R\rightarrow\right)  \simeq R^{2/3}$, which means that the model is not
consistence with the local gravity tests. \ 

\begin{figure}[ptb]
\includegraphics[height=7cm]{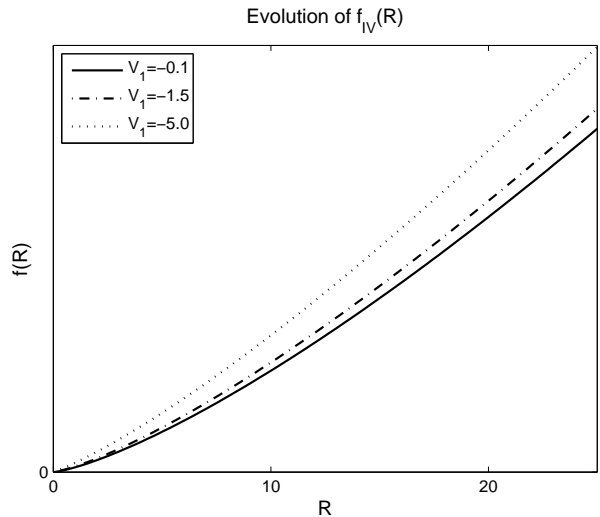}
%\mbox{\epsfxsize=14.2cm \epsffile{numericmodel2q.eps}}
\caption{Evolution of the function $f_{IV}\left(  R\right)  $, (\ref{fr.23}),
for different values of the constant $V_{1}$ in the range $R\in\left[
0,25\right]  $. For the plot we select $V_{2}=1$ with $f_{IV}\left(  0\right)
\simeq0$. \ The solid line is for $V_{1}=-0.1,$ the dash-dot line is for
$V_{1}=-1.5$ whereas the dot-dot line is for $V_{1}=-5.0$. \ From the plot it
is easy to see that holds $f_{IV}^{\prime}>0$,~and $f_{IV}^{\prime\prime}>0.$}%
\label{fiv}%
\end{figure}

In Fig. \ref{fiv}, we give the evolution of $f_{IV}\left(  R\right)  $ for
different values of $V_{1}$ in the same range for $R$. From the figure we
observe that, when $V_{1}$ increases, the evolution of $f_{IV}\left(
R\right)  $ is almost linear. Specifically the dot-dot line, which is for
$V_{2}=1,~V_{1}=-5$, $R\in\left[  0,25\right]  $, is approximated very well by
the quadratic polynomial $f\left(  R\right)  =a_{1}R+a_{2}R^{2}+a_{0},~$where
the constants, $a_{1-3}$, are\ $a_{1}\simeq3.7,~a_{2}\simeq8~10^{-3}~$and
$a_{3}\simeq-2.5~10^{-2}$. \ For higher-order polynomials of the form
$f\left(  R\right)  =%
%TCIMACRO{\dsum \limits_{K=0}}%
%BeginExpansion
{\displaystyle\sum\limits_{K=0}}
%EndExpansion
a_{K}R^{K}$,~we find that for $K>2,~\left\vert a_{K}\right\vert \lesssim
10^{-5}$. Furthermore we observe that $f_{IV}^{\prime}>0,~f_{IV}^{\prime
\prime}>0$, which indicates that model is ghost free. We note that the same
results hold for the $f_{V}\left(  R\right)  $ model.

The $f_{I}\left(  R\right)  ~$and $f_{II}\left(  R\right)  $ models for
$V_{1}=0,$~or $V_{1}\neq0,$ are not new and have previously been found in
\cite{palfr} from the application of Noether's Theorem for point
transformations to the Lagrangian, (\ref{fr.13}), of the field equations, and
belong to the family of models $f\left(  R\right)  =\left(  R^{b}%
-2\Lambda\right)  ^{c}$~~\cite{fr3}. The field equations for those models
admit Noether point symmetries which form\footnote{In the Mubarakzyanov
Classification Scheme
\cite{Morozov58a,Mubarakzyanov63a,Mubarakzyanov63b,Mubarakzyanov63c}} the
$A_{3}$ and the~$A_{3,8}~$(or $sl\left(  2,R\right)  $) Lie algebras,
respectively. Moreover $f_{II}\left(  R\right)  $ model is the Ermakov-Pinney
system, in $M^{2}$, and the quadratic conservation law, (\ref{fr.16}), is the
Ermakov-Lewis invariant. The closed-form solutions of $f_{I}\left(  R\right)
$ and $f_{II}\left(  R\right)  ~$models can be found in \cite{palfr,palfr2}.
For both models, for $V_{1}=0$ the closed-form solution for the scale factor
is a power law whereas for $V_{1}\neq0~$the scale factor has exponential
expansion in which the late-time solution describes the de Sitter Universe.

\subsection{Existence of de Sitter solutions}

\label{desitter}

In \cite{BarOtt}, it has been shown that for a flat FLRW spacetime, $f\left(
R\right)  $-gravity (without a matter source) provide de Sitter solutions,
that is, $R=R_{0}$, when the following expression holds%
\begin{equation}
R_{0}f^{\prime}\left(  R_{0}\right)  -2f\left(  R_{0}\right)  =0.
\label{fr.26a}%
\end{equation}
Note that from the power law models $f\left(  R\right)  =R^{n}$, only the
quadratic model, $n=2$, satisfies identically the condition (\ref{fr.26a}).
Moreover, since $f^{\prime}=\phi$, and $V_{,\phi}=R$, expression
(\ref{fr.26a}) can be written as follows
\begin{equation}
2V\left(  \phi_{0}\right)  -\phi_{0}V_{,\phi}|_{\phi=\phi_{0}}=0
\label{fr.26b}%
\end{equation}
where $\phi_{0}=f^{\prime}\left(  R_{0}\right)  $. We apply the last condition
to find de Sitter solutions for the models which followed from the application
of the group invariants.

For the potential $V_{I}\left(  \phi\right)  $, the application of
(\ref{fr.26b}) gives the that $\left(  \phi_{0}\right)  ^{2}=\left(
V_{1}/V_{2}\right)  $, while from the potential $V_{II}\left(  \phi\right)  ,$
we find that the only real solution $\left(  \phi_{0}\right)  ^{2}=\sqrt
{3}\left(  V_{2}/V_{1}\right)  ^{1/4}$. Furthermore for $V_{III}\left(
\phi\right)  $, from (\ref{fr.26b}) we have that it provides de Sitter
solution for $\phi_{0}=\frac{25}{16}\left(  V_{2}/V_{1}\right)  ^{2}$, whereas
the $V_{IV}\left(  \phi\right)  $-model gives a de Sitter solution for
$\phi_{0}=-\frac{1}{2}\left(  V_{1}/V_{2}\right)  $. Here we would like to
remark that condition (\ref{fr.26b}) for the $V_{I}\left(  \phi\right)  $ and
$V_{IV}\left(  \phi\right)  $ models holds and for $\phi_{0}=0$, however that
leads to not physical accepted solutions.

Finally for the $f_{V}\left(  R\right)  $-model, i.e. potential $V_{V}\left(
\phi\right)  $, the application of condition (\ref{fr.26b}) gives four points
which are%
\begin{equation}
\left(  \phi_{0}\right)  ^{2}=\beta~,~\phi_{0}=-\frac{V_{1}\mp\sqrt{V_{1}%
^{2}-16\beta\left(  V_{2}\right)  ^{2}}}{4V_{2}}.
\end{equation}

In the de Sitter solution, $a\left(  t\right)  =\exp\left(  H_{0}t\right)  $,
from (\ref{fr.05}) we have that $R_{0}=12H_{0}^{2}$. Hence by using the
relation $V_{,\phi}|_{\phi=\phi_{0}}=R_{0}$, we can derive the value of the
Ricciscalar. That provides us with information about the possible values of
the free parameters of the models.

For the $f_{I}\left(  R\right)  $, and $f_{II}\left(  R\right)  $ models, we
find that $R_{0}\simeq V_{1}$, which indicates that $V_{1}\simeq H_{0}^{2}$.
For the third model, namely $f_{III}\left(  R\right)  $, we find that
$R_{0}=\frac{32}{125}\left(  V_{1}\right)  ^{3}\left(  V_{2}\right)  ^{-2}$,
that is, $\left(  V_{1}\right)  ^{3}=\frac{375}{8}\left(  V_{2}H_{0}\right)
^{2}$, while for the $f_{IV}\left(  R\right)  $-model we have that in the de
Sitter point $\left(  V_{1}\right)  ^{3}=48\left(  V_{2}H_{0}\right)  ^{2}$.

Finally for the $f_{V}\left(  R\right)  $ model, which provides us with four
possible points with a de Sitter expansion the corresponding values of the
Ricciscalar are
\begin{equation}
R_{0}=4\beta\left(  V_{1}\mp4V_{2}\sqrt{\beta}\right)  , \label{fr.26c1}%
\end{equation}
for $\phi_{0}=\pm\sqrt{\beta}$, respectively and,
\begin{equation}
R_{0}=\frac{V_{1}}{8V_{2}^{2}}\left(  V_{1}^{2}-16V_{2}^{2}\beta\right)
\mp\frac{\left(  V_{1}^{2}-16V_{2}^{2}\beta\right)  ^{\frac{3}{2}}}{8V_{2}%
^{2}} \label{fr.26c}%
\end{equation}
for the last two points. Since $R_{0}>0$, in the de Sitter point, expressions
(\ref{fr.26c1}), (\ref{fr.26c}) provide us with constraints for the possible
physical accepted values of the free parameters of the model.

In the following Section we use the extra conservation laws in order to reduce
the order of the field equations for the three new integrable models,~$V_{III}%
\left(  \phi\right)  ,~V_{IV}\left(  \phi\right)  $ and $V_{V}\left(
\phi\right)  $. We do that by using the Hamilton-Jacobi theory.

\section{Solutions of the field equations}

\label{solutions}

In the coordinate system $\left\{  a,\phi\right\}  $, the Lagrangian
(\ref{fr.13}) of the field equations is as follows%
\begin{equation}
L\left(  a,\dot{a},\phi,\dot{\phi}\right)  =6a\phi\dot{a}^{2}+6a^{2}\dot
{a}\dot{\phi}+a^{3}V\left(  \phi\right)  . \label{fr.27}%
\end{equation}

From the last expression we define the momenta, $p_{a}=\frac{\partial
L}{\partial\dot{a}},~p_{\phi}=\frac{\partial L}{\partial\dot{\phi}}$, as
follows%
\begin{equation}
p_{a}=12a\phi\dot{a}+6a^{2}\dot{\phi}~,~p_{\phi}=6a^{2}\dot{a}. \label{fr.28}%
\end{equation}
The Hamiltonian, (\ref{fr.14}), in terms of the momentum has the following
expression
\begin{equation}
\mathcal{E}=\frac{1}{6a^{2}}\left(  p_{a}p_{\phi}-\frac{\phi}{a}p_{\phi}%
^{2}\right)  -a^{3}V\left(  \phi\right)  . \label{fr.30}%
\end{equation}

Moreover the field equations, (\ref{fr.05}) and (\ref{fr.07}), are equivalent
to the following Hamiltonian system%
\begin{equation}
\dot{a}=\frac{1}{6a^{2}}p_{\phi}~,~\dot{\phi}=\frac{1}{6a^{2}}p_{a}-\frac
{\phi}{3a^{3}}p_{\phi}, \label{fr.31}%
\end{equation}%
\begin{equation}
\dot{p}_{a}=\frac{p_{a}p_{\phi}}{3a^{3}}-\frac{\phi}{2a^{4}}p_{\phi}%
^{2}+3a^{2}V\left(  \phi\right)  \quad\mbox{\rm and} \label{fr.32}%
\end{equation}%
\begin{equation}
\dot{p}_{\phi}=\frac{p_{\phi}^{2}}{6a^{3}}+a^{3}V_{,\phi}. \label{fr.33}%
\end{equation}

From the Hamiltonian (\ref{fr.30}) we define the Hamilton-Jacobi Equation%
\begin{equation}
\frac{1}{6a^{2}}\left(  \left(  \frac{\partial S}{\partial a}\right)  \left(
\frac{\partial S}{\partial\phi}\right)  -\frac{\phi}{a}\left(  \frac{\partial
S}{\partial\phi}\right)  ^{2}\right)  -a^{3}V\left(  \phi\right)  -\left(
\frac{\partial S}{\partial t}\right)  =0 \label{fr.34}%
\end{equation}
in which $p_{a}=\frac{\partial S}{\partial a},$ $p_{\phi}=\frac{\partial
S}{\partial\phi}$, and $S=S\left(  t,a,\phi\right)  $. Equation (\ref{fr.34})
provides us with the action $S$, which help us to reduce the dimension of the
Hamiltonian system (\ref{fr.31}), (\ref{fr.32}). \ In the following we use the
classification of Darboux \cite{Darboux} for the integrable systems in a
two-dimensional manifold by following the notation of \cite{Daskaloy}.

Before we proceed, as a final remark we would like to express the EoS which
corresponds to the $f\left(  R\right)  $ terms in the coordinates,, $\left\{
a,\phi\right\}  $. From expression (\ref{fr.12a}) \ we have that%
\begin{equation}
w_{f}=-\frac{4H\dot{\phi}+2\ddot{\phi}-V\left(  \phi\right)  }{6H\dot{\phi
}-V\left(  \phi\right)  }. \label{fr.34a}%
\end{equation}
However, from (\ref{fr.27}) we calculate the \textquotedblleft
Klein-Gordon\textquotedblright\ equation for the field $\phi$, namely%
\begin{equation}
2\ddot{\phi}+4H\dot{\phi}-2\phi H^{2}-V\left(  \phi\right)  +\frac{1}{3}\phi
V_{,\phi}=0 \label{fr.34b}%
\end{equation}
by replacing $\ddot{\phi}$ in (\ref{fr.34a}) from (\ref{fr.34b}). We find%
\begin{equation}
w_{f}=-\frac{6\phi H^{2}-2\phi V_{,\phi}}{18H\dot{\phi}-3V\left(  \phi\right)
}, \label{fr.25b}%
\end{equation}
that is, the parameter in the EoS is expressed only in terms of the first
derivatives of $\left\{  a,\phi\right\}  $.

\subsection{$f_{III}\left(  R\right)  $-model}

For the $f_{III}\left(  R\right)  $-model with effective potential
$V_{eff}=a^{3}V_{III}\left(  \phi\right)  $, we define the new coordinates
\begin{equation}
a=\sqrt{u}~,~\phi=\frac{1}{6}\frac{v^{2}}{\sqrt{u}} \label{fr.35}%
\end{equation}
in which the Hamiltonian (\ref{fr.30}) becomes%
\begin{equation}
\mathcal{E}=\frac{p_{u}p_{v}}{v}-\left(  V_{1}u^{\frac{3}{2}}-\sqrt{6}%
V_{2}\frac{u^{\frac{7}{4}}}{v}\right)  . \label{fr.36}%
\end{equation}

In the new coordinates the conservation law (\ref{fr.17}) has the following
expression%
\begin{equation}
I_{III}=p_{v}^{2}-2\frac{u}{v}p_{u}p_{v}+\frac{6}{5}V_{1}u^{\frac{5}{2}%
}-2\sqrt{6}V_{2}\frac{u^{\frac{11}{4}}}{v}, \label{fr.37}%
\end{equation}
that is, the field equations form an integrable dynamical system where the
supporting manifold is a Lie surface \cite{Daskaloy,Kaln}. We recall that
another cosmological model, which is integrable and for which the supporting
manifold is a Lie surface, is a specific case of the early dark energy model
of a minimally coupled scalar field, for details see \cite{dynsym}.

Therefore we have that the action, $S$, in the new coordinates, $\left\{
u,v\right\}  $, has the following form%

\begin{equation}
S\left(  t,u,v\right)  =-\varepsilon\left(  vS_{0}\left(  u\right)  +\int
\frac{\sqrt{6}V_{2}u^{\frac{7}{4}}}{S_{0}\left(  u\right)  }du\right)
-\mathcal{E}t, \label{fr.27c}%
\end{equation}
where $S_{0}\left(  u\right)  =\sqrt{2\mathcal{E}u+I_{III}+\frac{4}{5}%
V_{1}u^{\frac{5}{2}}},$ and $\varepsilon=\pm1$.

With the use of (\ref{fr.27c}) the field equations are reduced to the
following two first-order ordinary differential equations%
\begin{equation}
v\dot{u}=-\varepsilon S_{0}\left(  u\right)  , \label{fr.38a}%
\end{equation}
and%
\begin{equation}
v\dot{v}=\varepsilon\left(  -v\frac{\mathcal{E}+V_{1}u^{\frac{3}{2}}}%
{S_{0}\left(  u\right)  }+\frac{\sqrt{6}V_{2}u^{\frac{7}{4}}}{S_{0}\left(
u\right)  }\right)  . \label{fr.39a}%
\end{equation}

Dynamical systems supported by a Lie surface cannot necessarily be solved by
the method of separation of variables. However, the importance of the
existence of the Lie surface is that we can solve the reduced system and
express the one dependent variable in terms of the other. \ From
(\ref{fr.38a}) we have $\frac{dv}{dt}=-\frac{\varepsilon S_{0}\left(
u\right)  }{v}\frac{dv}{du}.$ Hence equation (\ref{fr.39a}), becomes,%
\begin{equation}
\frac{dv}{du}=v\frac{\mathcal{E}+V_{1}u^{\frac{3}{2}}}{\left(  S_{0}\left(
u\right)  \right)  ^{2}}-\frac{\sqrt{6}V_{2}u^{\frac{3}{2}}}{\left(
S_{0}\left(  u\right)  \right)  ^{2}}. \label{fr.39b}%
\end{equation}

The solution of the latter is
\begin{equation}
v\left(  u\right)  =\left[  \int B\left(  u\right)  e^{-\int A\left(
u\right)  du}du+v_{0}\right]  e^{\int A\left(  u\right)  du}, \label{fr.40}%
\end{equation}
where%
\begin{equation}
A\left(  u\right)  =\frac{\mathcal{E}+V_{1}u^{\frac{3}{2}}}{\left(
S_{0}\left(  u\right)  \right)  ^{2}}~,~B\left(  u\right)  =-\frac{\sqrt
{6}V_{2}u^{\frac{7}{4}}}{\left(  S_{0}\left(  u\right)  \right)  ^{2}}.
\label{fr.41}%
\end{equation}

For instance, when $I_{3}=0$ and $\mathcal{E}=0$, the closed-form solution of
(\ref{fr.40}) in terms of the scale factor is\footnote{We note that for
$I_{3}=0$ and $V_{1}=0$ the closed-form solution of (\ref{fr.40}) is expressed
in terms of the Whittaker Function.}
\begin{equation}
v\left(  a\right)  =v_{1}a^{\frac{1}{2}}+v_{0}a^{\frac{5}{2}}, \label{fr.412}%
\end{equation}
where, $v_{1}=\frac{5\sqrt{6}}{4}\frac{V_{2}}{V_{1}}$. Hence from
(\ref{fr.35}) for the field, $\phi=f^{\prime}\left(  R\right)  ,$ we have the
following expression,
\begin{equation}
\phi\left(  a\right)  =\frac{1}{6}\left(  v_{1}+2v_{0}v_{1}a^{2}+v_{0}%
^{2}a^{4}\right)  .
\end{equation}

From (\ref{fr.35}) and (\ref{fr.38a}) the Hubble function, (\ref{fr.40}), is
\begin{equation}
H\left(  a\right)  =\frac{1}{2}\frac{S_{0}\left(  u\right)  }{u}\left[  \int
B\left(  u\right)  e^{-\int A\left(  u\right)  du}du+v_{0}\right]
^{-1}e^{-\int A\left(  u\right)  du},
\end{equation}
in which for the solution, (\ref{fr.412}), the Hubble function is%
\begin{equation}
H\left(  a\right)  =\sqrt{\frac{4}{5}V_{1}}a^{-\frac{1}{4}}\left(  v_{1}%
+v_{2}a^{2}\right)  ^{-1}.
\end{equation}

\begin{figure}[ptb]
\includegraphics[height=7cm]{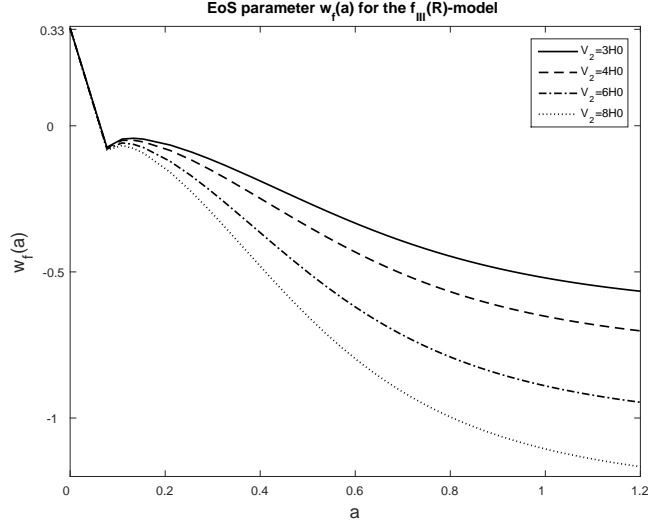}
%\mbox{\epsfxsize=14.2cm \epsffile{numericmodel2q.eps}}
\label{eos3}\caption{Evolution of the EoS parameter $w_{f}\left(  a\right)  $
for the $f_{III}\left(  R\right)  $-model. We observe that in the early
universe the fluid which follow from $f\left(  R\right)  $-gravity act like a
radiation fluid, i.e., $w_{f}\left(  a\right)  \simeq1/3~$ (see \cite{Ame10}).
Moreover at the late universe the EoS parameter can cross the phantom-barrier
and has a linear behavior which reach the deSitter point. For the plots we
select: $\mathcal{E}=6~\Omega_{m0}H_{0}^{2}$,~$I_{III}=0.2H_{0}^{2}$,
$V_{1}=H_{0}\left(  V_{2}H_{0}\right)  ^{\frac{2}{3}}$, $\varepsilon v\left(
0\right)  =0.15$, $\Omega_{m0}=0.28$, in units where $H_{0}=69.6~km/s/Mpc.$
The solid line is for $V_{2}=3H_{0},$ the dash-dash line is for $V_{2}=4H_{0}%
$, the dash-dot line is for $V_{2}=6H_{0}$, whereas the dot-dot line is for
$V_{2}=8H_{0}$.}%
\end{figure}

In order to study the behavior of the cosmological fluid, in fig. \ref{eos3}
we give the evolution for the EoS parameter (\ref{fr.25b}) which follows from
the solution (\ref{fr.40}) for the $f_{III}\left(  R\right)  $-model. We
observe that the effective perfect fluid which follows from the terms which
arise from the modified Friedmann's equations has an EoS parameter $w_{f}%
\leq\frac{1}{3}$ in which can cross the phantom-divide line, $w_{f}<-1$, in
the late universe for different values of the free parameters. The values of
the free parameters $V_{1},V_{2}$ has been chosen such as to approximate the
condition which follow from section \ref{desitter} and the solution to give a
de Sitter universe. Note that the Hubble constants which are provided by the
models are $H_{0}\simeq69.6km/s/Mpc$.

\subsection{$f_{IV}\left(  R\right)  $-model}

For the $f_{IV}\left(  R\right)  $-model we define the new variable $w=6a\phi
$. Hence the Hamiltonian, (\ref{fr.30}), and the quadratic conservation law,
(\ref{fr.18}), are written as follows,
\begin{equation}
\mathcal{E}=\frac{p_{a}p_{w}}{a}-\left(  \bar{V}_{1}w^{3}+\bar{V}_{2}%
\frac{w^{4}}{a}\right)  , \label{fr.43}%
\end{equation}%
\begin{equation}
I_{IV}=p_{a}^{2}-2\frac{w}{a}p_{a}p_{v}+\frac{3}{2}\bar{V}_{1}w^{4}+2\bar
{V}_{2}\frac{w^{5}}{a}, \label{fr.44}%
\end{equation}
where $M_{1}=6^{-3}V_{1}$ and $M_{2}=6^{-4}V_{2}$. Easily we can see that the
$f_{IV}$-model is integrable in which the supporting manifold is a Lie
surface, similarly to the $f_{III}$-model.

From (\ref{fr.43}) with the use of (\ref{fr.44}) we find that the solution of
the Hamilton-Jacobi equation is%
\begin{equation}
S\left(  t,a,w\right)  =\varepsilon\left(  \frac{a}{2}S_{1}\left(  w\right)
-2\int\frac{\bar{V}_{2}w^{4}}{S_{1}\left(  w\right)  }\right)  -\mathcal{E}t,
\end{equation}
where $\varepsilon=\pm1$~and $S_{1}\left(  w\right)  =\sqrt{2M_{1}%
w^{4}+8\mathcal{E}w+4I_{IV}}$

Therefore the field equations are reduced to the following system
\begin{equation}
a\dot{a}=-2\varepsilon\frac{a\left(  \mathcal{E}+M_{1}w^{3}\right)
+M_{2}w^{4}}{S_{1}\left(  w\right)  }, \label{fr.46}%
\end{equation}
and%
\begin{equation}
a\dot{w}=-\frac{\varepsilon}{2}S_{1}\left(  w\right)  . \label{fr.47}%
\end{equation}

Moreover in the limit for which, $\mathcal{E}=0$,~~$I_{IV}=0,$ and~$M_{2}=0,$
the closed-form solution of the scale factor is $a\left(  t\right)
=a_{0}t^{2}$, in which we have applied the initial condition $a\left(
t\rightarrow0\right)  =0$. That solution corresponds to $f_{IV}^{1}\left(
R\right)  \simeq R^{3}~$model and we can see that the solution describes a
universe with a perfect fluid in which the parameter in the EoS is
$w_{f}=-\frac{1}{3}$.

Similarly with the $f_{III}\left(  R\right)  $-model we can solve the one
dependent parameter of the system (\ref{fr.46}), (\ref{fr.47}) in terms of the
other. However, for the $f_{III}\left(  R\right)  $-model we did that by
expressing the new variable, $v$, in terms of the scale factor. In this model
we can express the scale factor in terms of the new variable, $w,$ i.e.
$a\left(  w\right)  $.

Hence with the use of (\ref{fr.47}) equation (\ref{fr.46}) becomes
\begin{equation}
\frac{da}{dw}=\bar{A}\left(  w\right)  a+\bar{B}\left(  w\right)  ,
\label{fr.48}%
\end{equation}
where
\begin{equation}
\bar{A}\left(  w\right)  =\frac{4\left(  \mathcal{E}+M_{1}w^{3}\right)
}{\left(  S_{1}\left(  w\right)  \right)  ^{2}},~\bar{B}\left(  w\right)
=\frac{4M_{2}w^{4}}{\left(  S_{1}\left(  w\right)  \right)  ^{2}},
\label{fr.49}%
\end{equation}
that is, the solution of $a\left(  w\right)  $ is given by the formula
(\ref{fr.40}).

We continue to the analysis of the last integrable model, namely $f_{V}\left(
R\right)  $-model.

\subsection{$f_{V}\left(  R\right)  $-model}

For the last model which is given by the application of Killing tensors,
contact symmetries, to the Lagrangian of the field equations we can see that
the supporting manifold of the dynamical system is a Liouville surface. Hence
the Hamilton-Jacobi Equation can be solved with the method of separation of variables.

Under the coordinate transformation%
\begin{equation}
a=x+y~,~\phi=\sqrt{\beta}\frac{x-y}{x+y} \label{fr.50}%
\end{equation}
the Hamiltonian function (\ref{fr.30}) is
\begin{equation}
\mathcal{E=}\frac{\frac{1}{2}p_{x}^{2}-\frac{1}{2}p_{y}^{2}-U_{1}x^{4}%
-U_{2}y^{4}}{12\sqrt{\beta}\left(  x+y\right)  }, \label{fr.51}%
\end{equation}
where the constants $U_{1}$ and $U_{2}$, are $U_{1}=12\beta^{2}\left(
V_{1}+4V_{2}\sqrt{\beta}\right)  $ and $U_{2}=-12\beta^{2}\left(  V_{1}%
-4V_{2}\sqrt{\beta}\right)  $. \ We see that, when $V_{1}^{2}=16\beta
V_{2}^{2}$, we have that $U_{1}=0$ or $U_{2}=0$.

Furthermore the quadratic conversation law (\ref{fr.19}) in the new
coordinates has the following form%
\begin{equation}
I_{V}=\frac{yp_{x}^{2}+xp_{y}^{2}-2\left(  U_{1}yx^{4}-U_{2}xy^{4}\right)
}{x+y}. \label{fr.52}%
\end{equation}

Therefore the solution of the Hamilton-Jacobi Equation is
\begin{align}
S\left(  t,x,y\right)   &  =\varepsilon_{1}\int\sqrt{I_{5}-24\sqrt{\beta
}\mathcal{E}x+2U_{1}x^{4}}dx\nonumber\\
+  &  \varepsilon_{2}\int\sqrt{I_{5}+24\sqrt{\beta}\mathcal{E}y-2U_{2}y^{4}%
}dy, \label{fr.53}%
\end{align}
whereas the field equations reduce to the following system of first-order
ordinary differential equations%
\begin{equation}
12\sqrt{\beta}\left(  x+y\right)  \dot{x}=\varepsilon_{1}\sqrt{I_{5}%
-24\sqrt{\beta}\mathcal{E}x+2U_{1}x^{4}}, \label{fr.54}%
\end{equation}
and%
\begin{equation}
12\sqrt{\beta}\left(  x+y\right)  \dot{y}=-\varepsilon_{2}\sqrt{I_{5}%
+24\sqrt{\beta}\mathcal{E}y-2U_{2}y^{4}}, \label{fr.55}%
\end{equation}
where $\varepsilon_{1,2}=\pm1$.

Under the change of variable, $dt=\frac{1}{a}d\tau$, the closed-form solution
of the system (\ref{fr.54}), (\ref{fr.55}) is given in terms of Elliptic functions.

From (\ref{fr.50}) with the use of (\ref{fr.54}) and (\ref{fr.55}) we derive
the Hubble function%
\begin{align}
12\sqrt{\beta}H\left(  a\right)   &  =\varepsilon_{1}\sqrt{I_{5}-24\sqrt
{\beta}\mathcal{E}x+2U_{1}x^{4}}+\label{fr.56}\\
&  -\varepsilon_{2}\sqrt{I_{5}+24\sqrt{\beta}\mathcal{E}y-2U_{2}y^{4}},
\end{align}
In the limit for which $\dot{x}\simeq0,$ $x+y\simeq y,~$ $\bar{U}_{2}<0$, that
is,~$V_{1}>4V_{2}\sqrt{\beta}$, the latter can takes the following form%
\begin{equation}
\left(  \frac{H\left(  a\right)  }{H_{0}}\right)  ^{2}\simeq\Omega_{r0}%
a^{-4}+\Omega_{m0}a^{-3}+\Omega_{\Lambda} \label{fr.57}%
\end{equation}
in which $I_{5}=144\beta\Omega_{r0}H_{0}^{2}$, and $\Omega_{\Lambda}\simeq
U_{2}$. The last equation describe a universe in General Relativity with
cosmological constant, dark matter and radiation fluid. We can see that the
density of the radiation term, which is provided by the $f\left(  R\right)
$-theory is related with the value of the Noetherian conservation law $I_{5}$.
\ However this is only a particular solution and the general behavior of the
Hubble function is different.

From the system (\ref{fr.54}), (\ref{fr.55}) we observe that the free
parameters of the model that we have to determine are, $\mathcal{E}%
=6\Omega_{m0}H_{0}^{2}$, $\beta,$ $V_{1},~V_{2}$, the value of the
conservation law $I_{5}$, the initial conditions $\left(  x_{0},y_{0}\right)
=\left(  x,y\right)  |_{t\rightarrow t_{0}}$, and $\varepsilon_{1}%
,~\varepsilon_{2}$. In order to reduce the number of the free parameters we
apply the initial condition $a\left(  t\rightarrow0\right)  \simeq0^{+}$,
which gives that $x_{0}\simeq y_{0}$. Moreover in the de Sitter points, in
which the Ricciscalar is given by the expressions (\ref{fr.26c1}) and
(\ref{fr.26c}), we observe that if we set, $\beta\simeq H_{0}^{2}$, $\beta
V_{1}\simeq H_{0}^{2}$, and $\beta V_{2}\simeq H_{0},~$then $R_{0}\simeq
H_{0}^{2}$. \ As we discussed the value of the conservation law $I_{5}$ can be
related with the energy density of the radiation fluid which is introduced by
the theory, in the present era $\Omega_{r}$ is small, that is, we can assume
that $I_{5}\simeq0$. \ 

Furthermore we select that $\beta=H_{0}^{2},~V_{2}=\gamma H_{0}^{-1}$ and
$V_{1}=4\alpha\gamma$, with $\alpha>1,\gamma>0$, and the initial conditions
$\left(  x_{0},y_{0}\right)  $ are that in order the present value of the
Hubble constant to be $H_{0}=69.6km/s/Mpc~$\cite{BennetH0}. Finally we choose
the solution in which $\varepsilon_{1,2}=+1$, and the free parameters of the
problem to be $\left\{  \alpha,\gamma,\Omega_{m0}\right\}  $. \ 

\begin{figure}[ptb]
\includegraphics[height=7cm]{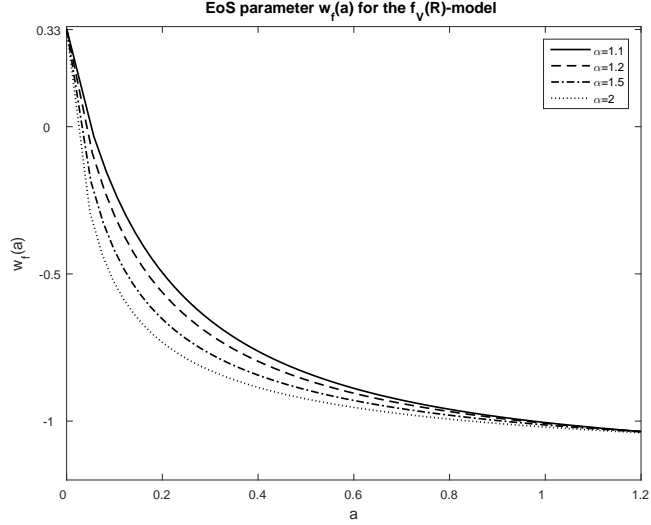}
%\mbox{\epsfxsize=14.2cm \epsffile{numericmodel2q.eps}}
\caption{Evolution of the parameter in the EoS, $w_{f}\left(  a\right)  $, for
the $f_{V}\left(  R\right)  $-model. For the numerical solutions we select
$\mathcal{E}=6\Omega_{m0}H_{0}^{2},~I_{IV}=0$, $V_{2}=H_{0}^{-1}$%
,~$V_{1}=4\alpha,$ $\varepsilon_{1,2}=+1$, $a\left(  0\right)  =0$ and
$\Omega_{m0}=0.28$ and $H_{0}=69.6~Km/s/Mpc.$ The solid line is for $a=1.1,$
the dash-dash line is for $\alpha=1.2$, the dash-dot line is for $\alpha=1.5$
and the dot-dot line is for $\alpha=2$. }%
\label{eos5a}%
\end{figure}

For $\gamma=1$, $\Omega_{m0}=0.28$, and for $\alpha\in\left(  1,2\right)  $,
in \ Fig. \ref{eos5a} the numerical evolution for the equation of state
parameter $w_{f}$, for the $f_{V}\left(  R\right)  $-model is given. From the
Fig. we observe that $w_{f}\leq1/3$ and $w_{f}$ can cross the phantom barrier.
Specifically, the EoS parameter $w_{f}~,$\ it decreases rapidly and takes a
negative value. Then the rate of decrease becomes slower where it has a linear
behavior of the form $w_{f}\left(  a\right)  =w_{1}a+w_{0},~w_{1}<0~,$ which
is the CPL parametric dark energy model introduced by Chevallier, Polarski
\cite{ch}, and Linder \cite{linder}.

\begin{figure}[ptb]
\includegraphics[height=7cm]{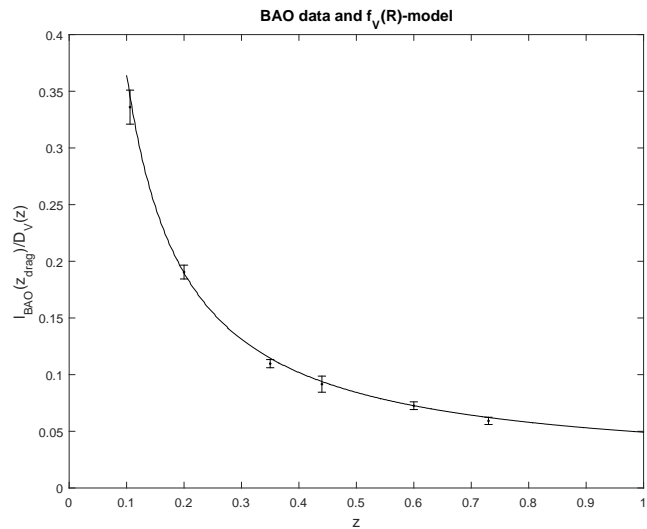}
%\mbox{\epsfxsize=14.2cm \epsffile{numericmodel2q.eps}}
\caption{BAO\ data and evolution of the $d_{th}=l_{BAO}\left(  z_{drag}%
\right)  \left(  D_{V}\left(  V\right)  \right)  ^{-1}$, parameter for the
$f_{V}\left(  R\right)  $-model for the parameters $\left(  \alpha
,\gamma,\Omega_{m0}\right)  _{fit}=\left(  1.24,0.8,0.28\right)  $ \ in which
$\left(  \chi_{\min}^{2}\right)  _{total}^{f_{V}}\simeq559.$}%
\label{baodata}%
\end{figure}

In order to test the viability of the $f_{V}\left(  R\right)  $-model we
perform a joint likelihood analysis using the Type Ia supernova data set of
Union 2.1 \cite{Suzuki}, and the BAO data \cite{Percival,BlakeC} in which we
select the free parameters of the model to be $\left\{  \alpha,\gamma
,\Omega_{m0}\right\}  .$ We fit the model with the data using the
gradient-search method \cite{robi}, for different set of random numbers in the
space of the free parameters in order to avoid local minimum in the chi-square
space, for the free parameters we select the range $\alpha\in\left(
1,2\right)  $, $\gamma\in\left(  0.5,1.2\right)  $, and $\Omega_{m0}=\left(
0.25,0.35\right)  $. \ We find that the best fit parameters are $\left(
\alpha,\gamma,\Omega_{m0}\right)  _{fit}=\left(  1.24,0.8,0.28\right)  $, in
which the $\left(  \chi_{\min}^{2}\right)  _{total}^{f_{V}}=\min\left(
\chi_{SNIa}^{2}+\chi_{BAO}^{2}\right)  \simeq559$, while with the same
algorithm, for the $\Lambda$-cosmology\footnote{Recall that in $\Lambda
$-cosmology, the Hubble function is $H\left(  z\right)  =H_{0}\left[  \left(
1-\Omega_{m0}\right)  +\Omega_{m0}\left(  1+z\right)  ^{3}\right]  ^{1/2}.$}
we find that $\left(  \chi_{\min}^{2}\right)  _{total}^{\Lambda}=\min\left(
\chi_{SNIa}^{2}+\chi_{BAO}^{2}\right)  \simeq560$. The small difference on the
minimum chi-square value between the two models, $\left\vert \Delta\chi_{\min
}^{2}\right\vert \simeq1$, indicates that the two models fit the data with a
similar way. \ In Fig. \ref{baodata} we plot the theoretical
parameter\footnote{Where $l_{BAO}\left(  z_{drag}\right)  $ is the BAO scale
at the drag redshift, and $D_{V}\left(  z\right)  $ is the volume distance}
$d_{zth}=l_{BAO}\left(  z_{drag}\right)  \left(  D_{V}\left(  z\right)
\right)  ^{-1}$, for the $f_{V}\left(  R\right)  $-model and the six BAO data
with the corresponding errors \cite{BlakeC}.

\section{Conclusions}

\label{con}

In this work we considered an FLRW spacetime in which the
gravitational\ Action Integral is that of $f\left(  R\right)  $-gravity with a
dustlike fluid. In order to determine the functional form of the $f\left(
R\right)  $ function we applied as selection rule the existence of Noether
symmetries for the field equations which are followed by Lie-B\"{a}cklund
transformations linear in the momentum, contact transformations. The
importance of this kind of symmetries is that they provide us with quadratic
conservation laws.

As the Lagrangian of the field equations is in the form of classical physical
dynamical systems of the form $L\left(  x^{k},\dot{x}^{k}\right)  =T-V$, where
$T$ is the kinetic energy and $V$ the potential, we were able to apply the
existence results in the literature in order to perform our classification.
Hence contact symmetries if $f\left(  R\right)  $-gravity is generated by the
Killing tensors of the minisuperspace, that is, the theory. In fact the
Killing tensors of the minisuperspace select the model.

For a spatially flat universe we found five models which admit quadratic
conservation laws, where for spatially nonflat universe we found only three
models which are included in the five models of the spatially flat FLRW
spacetime. From the five models, two models are well known in the literature
and they have been found from the application of Noether's Theorem for point
transformations. Moreover the quadratic conservation laws which correspond to
the five models are in involution with the Hamiltonian function, that is, the
field equations are integrable.

For the three new models with the use of the extra quadratic conservation laws
we reduced the field equations to a system of two nonlinear first-order
ordinary differential equations. We performed numerical simulations for the
models and we studied the evolution of the parameter in the EoS for the fluid
components which corresponds to the $f\left(  R\right)  $-theory. For all the
models we show that the parameter in the EoS has an upper bound which is
$w_{f}\leq1/3$. There is no lower bound which means that $w_{f}$ can cross the
phantom divide line. Furthermore, for the $f_{III}\left(  R\right)  $ and the
$f_{V}\left(  R\right)  $ models the cosmological fluid which follows from the
additional terms of the Friedmann equations in the present time the parameter
in the EoS has a linear behavior given by the linear function $w_{f}\left(
a\right)  =w_{1}a+w_{0}$. Furthermore, for the $f_{V}\left(  R\right)
$-model, and from (\ref{fr.56}), we showed that the value of the second
conservation law is related to the fluid components which are introduced by
$f\left(  R\right)  $-gravity and, specifically, it is related to the density
of the radiation fluid. That is an important result which indicates a relation
between the conservation laws and physical observable quantities. However, the
exact physical properties of the conservation laws of the field equations are
still unknown. Furthermore for the $f_{V}\left(  R\right)  $-model we showed
that can fit the cosmological data\ with a similar way with the $\Lambda-$cosmology.

Another issue that we did not discuss in this work is the connection of these
$f\left(  R\right)  $-models with other conformally equivalent theories. The
reason for which we restricted our analysis is because we considered the
dustlike fluid which does not interacting with gravity. Of course if we
relaxed that restriction or there is no dustlike fluid, i.e., $\mathcal{E}=0$,
then the solutions we have found also hold and for conformally equivalent
theories. However, in that case someone should extend the application of
contact transformations which leave invariant the field equations not only to
that which follow from the Killing tensors but also to the Conformal Killing
tensors. For a discussion on the relation of symmetries and conservation laws
of conformal equivalence theories see \cite{mgrg}.

This work extends the analysis of group invariant transformations in
gravitational physics and cosmology and shows that the application of group
invariants in modified theories provides us with models which can describe the
late-time acceleration phase of the universe.

\begin{acknowledgments}
The author thanks Professor PGL Leach for a useful discussion on the
application of contact symmetries in physical science, Professor F. Garufi for
fruitful discussion on the statistical analysis, Dr. S Pan for useful comments
and suggestions, and Professor M. Tsamparlis for a careful reading of the
manuscript and useful comments which improved the quality of the present work.
The work was supported by FONDECYT postdoctoral grant no. 3160121.
\end{acknowledgments}

%

%TCIMACRO{\TeXButton{appendix}{\appendix}}%
%BeginExpansion
\appendix
%EndExpansion

\section{Quadratic conservation laws in spatially nonflat $f\left(  R\right)
$-models}

\label{nonflat}

In this Appendix we complete our analysis on the $f\left(  R\right)  $-models
which admit quadratic conservation laws if FLRW spacetime has nonvanishing
spatial curvature. Above we considered that the FLRW had zero spatial
curvature. In the case where the spatial curvature is $K,~$with $K\neq0$, from
the Action Integral, (\ref{fr.02}), and with the use of the Lagrange
Multiplier we find the following Lagrangian for the field equations
\begin{equation}
L\left(  a,\dot{a},\phi,\dot{\phi}\right)  =6a\phi\dot{a}^{2}+6a^{2}\dot
{a}\dot{\phi}+a^{3}V\left(  \phi\right)  -6Ka\phi. \label{nf.01}%
\end{equation}

We can see that this Lagrangian admits the same minisuperspace as the
Lagrangian, (\ref{fr.27}). Therefore in order to apply the method of Section
(\ref{LBsym}) for the existence of a contact transformation which leaves the
Action Integral, (\ref{fr.02}), invariant with Lagrangian, (\ref{nf.01}), we
use the KTs of Appendix \ref{killing}.

Hence, when $V\left(  \phi\right)  =V_{I}\left(  \phi\right)  $, the modified
field equations admit the conservation law
\begin{equation}
\bar{I}_{I}=I_{I}\left(  a,\dot{a},\phi,\dot{\phi}\right)  \text{.}
\label{nf.02}%
\end{equation}

Moreover, in the case for which $V\left(  \phi\right)  =V_{IV}\left(
\phi\right)  $, the quadratic conservation law of the field equations is%
\begin{equation}
\bar{I}_{IV}=I_{IV}\left(  a,\dot{a},\phi,\dot{\phi}\right)  -6K\left(
\alpha\phi\right)  ^{2}, \label{nf.03}%
\end{equation}
whereas for $V_{V}\left(  \phi\right)  $ the quadratic conservation law is
\begin{equation}
\bar{I}_{V}=I_{V}\left(  a,\dot{a},\phi,\dot{\phi}\right)  -3Ka^{2}\left(
\phi^{2}-\beta\right)  . \label{nf.04}%
\end{equation}

Functions $I_{I},~I_{IV}$ and $I_{IV}$ are given by the expressions
(\ref{fr.15}), (\ref{fr.18}) and (\ref{fr.19}), respectively.

\section{Killing vectors and Killing tensors}

\label{killing}

The minisuperspace, (\ref{tr.10}), is the $M^{2}$ space, which means that it
admits a three-dimensional Killing algebra. The KVs in the coordinates
$\left\{  a,\phi\right\}  $ are:
\begin{equation}
K_{1}^{i}=a\partial_{a}-3\phi\partial_{\phi}~,~K_{3}^{i}=\frac{1}{a}%
\partial_{a}\quad\mbox{\rm and}
\end{equation}%
\begin{equation}
K_{2}^{i}=\frac{1}{a}\left(  \partial_{a}-\frac{\phi}{a}\right)  \partial_{a}.
\end{equation}

Moreover, $M^{2},$ admits five KTs (except the metric tensor $\gamma_{ij}$),
which are of the form $K_{AB}^{ij}=K_{A}^{(i}\otimes K_{B}^{j)}$, where
$A,B=1,2,3$. Hence the five KTs in the coordinates $\left\{  a,\phi\right\}  $
are%
\begin{equation}
K_{11}^{ij}=%
\begin{pmatrix}
a^{2} & -3a\phi\\
-3a\phi & 9\phi^{2}%
\end{pmatrix}
~,~K_{22}^{ij}=%
\begin{pmatrix}
1/a^{2} & -\phi/a^{3}\\
-\phi/a^{3} & \phi^{2}/a^{4}%
\end{pmatrix}
,
\end{equation}

\begin{equation}
K_{33}^{ij}=%
\begin{pmatrix}
0 & 0\\
0 & a^{-2}%
\end{pmatrix}
~,~K_{12}^{ij}=%
\begin{pmatrix}
1 & -2a/\phi\\
-2a/\phi & 3\phi^{2}/a^{2}%
\end{pmatrix}
~ \mbox{\rm and}
\end{equation}
and%
\begin{equation}
K_{13}^{ij}=\frac{1}{2}%
\begin{pmatrix}
0 & 1\\
1 & -6\phi/a^{2}%
\end{pmatrix}
.
\end{equation}

Recall that $K_{23}^{ij}=K_{2}^{(i}\otimes K_{3}^{j)}$ is the metric tensor
$\gamma^{ij}$.

\end{document}